\documentstyle[twoside,fleqn,espcrc2]{article}

\newcommand{\AmS}{{\protect\the\textfont2
  A\kern-.1667em\lower.5ex\hbox{M}\kern-.125emS}}

\newpage

\title{The Interface Tension in Quenched QCD at the Critical
Temperature\thanks{Talk presented by Bernd Grossmann }}
\author{B.~Grossmann\address{HLRZ, c/o Kfa Juelich, P.O.~Box 1913,
                             D-5170 J\"ulich, Germany},
        M.~L.~Laursen$^{\rm a}$,
        T.~Trappenberg$^{\rm a\,}$
            \address{Institute for Theoretical Physics E,
            RWTH Aachen, \\Sommerfeldstr.,D-5100 Aachen, Germany}
            \thanks{Supported by Deutsche Forschungsgemeinschaft}
        and
        U.~J.~Wiese\address{Universit\"at Bern, Sidlerstr. 5, 3012 Bern,
                            Switzerland}
            \thanks{Supported by Schweizer Nationalfond}   }
\begin{document}
\begin{abstract}
We present results for the confinement-deconfinement interface tension
$\alpha_{cd}$ of quenched QCD. They were obtained by applying Binder's
histogram method to lattices of size $L^2\times L_z\times L_t$ for $L_t=2$ and
$L=8,10,12\mbox{ and }14$ with $L_z=30$ for $L=8$ and $L_z=3L$ otherwise.
The use of a multicanonical algorithm and cylindrical geometries have turned
out to be crucial for the numerical studies.
\end{abstract}

\maketitle

\section{Introduction}

At high temperature a phase transition occurs in QCD. In the quenched
approximation (i.e. without any light quarks) this transition is of first order
and separates a low temperature confined phase from a high temperature
deconfined phase. The dynamics of a system which crosses the transition
temperature $T_c$ (as e.g. in the early universe or in heavy ion collisions)
depends on the free energy
\begin{equation}
F_{cd} = \alpha_{cd}\, A
\end{equation}
of an interface of area $A$ between regions of confined and deconfined matter.
The interface tension $\alpha_{cd}=\sigma_{cd}\, T_c$ was investigated before
in Monte Carlo simulations of lattice systems with $L_t=2$ using various
approaches (see \cite{kaj91:hist,hua90:hist,bro92:hist,gro92b:histesp}).
Lately these results have been questioned
based on an application of Binder's histogram method to cubic spatial volumes
$L^3$ with $L=6,8,10,\mbox{ and } 12$ (see \cite{jan92:hist,gro92:histas}).
However, these results might have been plagued by interfacial interactions.
Therefore, we present results using the same method but on asymmetric volumes
($L_z>L_x=L_y$) thereby reducing these interactions.

\section{The Interfacial Free Energy}
\label{interfacial}
We consider SU(3) pure gauge theory with the Wilson action $S$ on a cylindrical
lattice of size $L_x\times L_y\times L_z\times L_t$ with $L_z=L_y=L\mbox{ and }
L_z\ge 3L$ at the critical
coupling $\beta_c$ for $L_t=2$. We use periodic boundary conditions in the
time direction and $C-$periodic boundary conditions \cite{kro91:hist} in
the spatial directions, i.e.
\begin{eqnarray}
U_\mu(\vec{x} + L_i \vec{e}_i,t) &=& U_\mu^*(\vec{x},t),\mbox{ for }i=x,y,z\\
U_\mu(\vec{x} ,t + L_t) & = & U_\mu(\vec{x},t).
\end{eqnarray}
Because of the $C-$periodic boundary conditions the value of the Polyakov line
$\Omega_L(\vec{x}) \equiv tr\left(\prod_{t=1}^{L_t}U_0(\vec{x},t)\right)$
will satisfy
\begin{equation}
\Omega_L(\vec{x}+ L_i\vec{e}_i) = \Omega_L^*(\vec{x})\mbox{ for }i=x,y,z
\end{equation}
Therefore, no bulk configurations in either of the two deconfined phases that
have nonvanishing imaginary part of
$\Omega_L\equiv 1/(L^2L_z)\sum_{\vec{x}} \Omega_L(\vec{x})$
can exist and the probability distribution $P_L(\rho)d\rho$ of
$\rho\equiv Re\,\Omega_L$ takes the form sketched in Fig.\ref{fig:pmc}.
\begin{figure}[t]
\vspace*{-0.9cm}
\vspace{6cm}
\caption{ {
Schematic probability distribution for the order parameter. The dotted
line indicates the multicanonical distribution.
}}
\label{fig:pmc}
\end{figure}
The system is most likely in either the one remaining
deconfined phase corresponding to $\rho^{(2)}$ or the confined phase
at $\rho^{(1)}$. When $\rho$ is increased from $\rho^{(1)}$,
bubbles of deconfined phase form. These
configurations are suppressed by the interfacial free energy
$\sigma_{cd} A$, where  $A$ is the surface area of the bubble. It grows
until finally its surface is larger than the surface $L^2$ of two planar
interfaces which devide the lattice into three parts as depicted in the second
part of Fig.~\ref{fig:config}.
\begin{figure}[t]
\vspace{4cm}
\caption{ {
Typical cuts through the lattice in the $y-z-$plane at values $\rho^{(1)}$
(first picture) and $\rho^{(min)}$ (second picture) of the order parameter. The
two phases are represented by white resp. shaded areas.
}}
\label{fig:config}
\end{figure}
Since the interface area of the two planar interfaces is independent of
$\rho$ the probability $P_L$ is constant in the region where their
contributions dominate, i.e. around $\rho^{(min)}$.
Because of the $C-$periodic boundary conditions these interfaces
always separate a region in the confined phase from one in the deconfined
phase that has $Im(\Omega_L)=0$. Thus the corresponding configurations will be
exponentially suppressed by the interfacial free energy of two
confined-deconfined interfaces. Taking into account the capillary wave
fluctuations of the interfaces as well as their translational degrees of
freedom leads to additional power law corrections
\cite{bun92:histesp,wie92:histesp} giving
\begin{equation}
P_L^{min}\propto L_z^2\cdot L^{d-3}\cdot\exp\left( -2\sigma_{cd}L^{d-1}\right).
\label{eq:capwav}
\end{equation}
for $d-$dimensional spatial volumes.
This relation will be used to determine $\sigma_{cd}$ from the distributions
obtained on finite lattices.

In order to calculate the probability distribution $P_L(\rho)$, one has to
simulate the SU(3) pure gauge theory at the deconfinement phase transition. But
because of eq.~\ref{eq:capwav} any standard local updating algorithm will have
autocorrelation times $\tau_L$ which increase exponentially with $L^2$
("supercritical slowing down"). The use of the multicanonical algorithm
reduces this effect considerably.

\section{The Multicanonical Algorithm}
\label{multicanonical}

In order to overcome the supercritical slowing down, the multicanonical
algorithm \cite{ber91c:hist,mar92:histesp} does not sample
the configurations with the canonical Boltzmann weight
\begin{equation}
{\cal P}_L^{can}(S) \propto \exp (\beta S),
\end{equation}
where $S=1/3\,\sum_\Box\,tr\,U_\Box$ is the Wilson action in four dimensions,
but rather with a modified weight
\begin{equation}
{\cal P}_L^{mc}(S) \propto \exp (\beta_L(S) S+\alpha_L(S)).
\label{eq:boltzmc}
\end{equation}
The coefficients $\alpha_L$ and $\beta_L$ are chosen such that the
probability $P_L$ (not to be confused with the Boltzmann weights) is
increased
for all values of the action corresponding to the region between
$\rho^{(1)}\mbox{ and }\rho^{(2)}$, as shown schematically in
Fig.~\ref{fig:pmc}.
Details are described in \cite{gro92:histas} where the efficiency of the
multicanonical algorithm for SU(3) pure gauge theory has been demonstrated.
We apply the algorithm to the determination of the interfacial free energy.

\section{Numerical Results}
\label{results}

We have determined the probability distributions for $L_t=2$ and the spatial
volumes $L^2\times L_z$ with $L=8,\,10,\,12,\mbox{ and }14$ and $L_z=30$ for
$L=8$ and $L_z=3\, L$ otherwise.
Fig.~\ref{fig:interface} shows the real part of
$\Omega_L(z)\equiv 1/L^2\,\sum_{x,y}\Omega_L(x,y,z)$ for a typical
configuration close to $\rho^{(min)}$ on  a $14^2\times 42\times 2$
lattice.
\begin{figure}[t]
\vspace{6cm}
\caption{ {
$Re\,\Omega_L(z)$ for a $14^2\times 42 \times 2 $ lattice. The dotted lines
indicate the bulk expectation values of the two phases.
}}
\label{fig:interface}
\end{figure}
As expected from section \ref{interfacial}, one can
identify two interfaces between the confined phase and the deconfined phase.
The imaginary part of $\Omega_L$ is always zero. In
Fig.~\ref{fig:distribution} the resulting probability distributions are shown.
\begin{figure}[t]
\vspace{6cm}
\caption{ {
Measured probability distribution for the order parameter.
}}
\label{fig:distribution}
\end{figure}
In contrast to the distributions for cubic volumes  ($L$ values as before) they
all have a region of
constant probability in between the two peaks. This supports the
scenario developped in section~\ref{interfacial}.

In order to extract the interface tension we evaluate the quantities
\begin{equation}
 F_L^{(1)} \equiv \frac{1}{2 L^{2}}
\ln\frac{\overline{P_L}^{max}}{P_L^{min}}
            +\frac{3}{4}\frac{\ln L_z}{L^2}
            - \frac{1}{2}\frac{\ln L}{L^2}
 \end{equation}
and
\begin{equation}
 F_L^{(2)} \equiv -\frac{1}{2 L^{2}} \ln P_L^{min}
            + \frac{\ln L_z}{L^2}
\end{equation}
 where $\overline{P_L}^{max}\equiv \frac{1}{2}(P_L^{max,1}+P_L^{max,2} )$.
Note that one expects $\overline{P_L}^{(max)} \propto\sqrt{L_zL^2}$.
According to eq.~\ref{eq:capwav} both quantities should be linear functions of
$1/L^2$. Their intercept with the $y$-axis is  $\sigma_{cd}$.
We extract $F_L^{(1)}$ and $F_L^{(2)}$ from the  probability distributions of
Fig.~\ref{fig:distribution}. The results are plotted in
 Fig.~\ref{fig:fl} together with the corresponding linear fits. For the
interface tension we get the value
\begin{equation}
\frac{\alpha_{cd}}{T_c^3} = 0.10(1)\, .
\end{equation}
It agrees within errors with the value obtained by replacing the
Polyakov line by the action density. The agreement with
\cite{kaj91:hist,hua90:hist,bro92:hist} is good while  \cite{gro92b:histesp}
quotes a slightly higher value. Still the  discrepency between these results
and \cite{jan92:hist,gro92:histas} which used Binder's histogram method for
cubic volumes is reduced considerably  and can thus be attributed mainly
to interfacial interactions.
%
\begin{figure*}[t]
\vspace{6cm}
\caption{ {
Results for $F_L^{(1)}$ and $F_L^{(2)}$.
}}
\label{fig:fl}
\end{figure*}
%
%

%
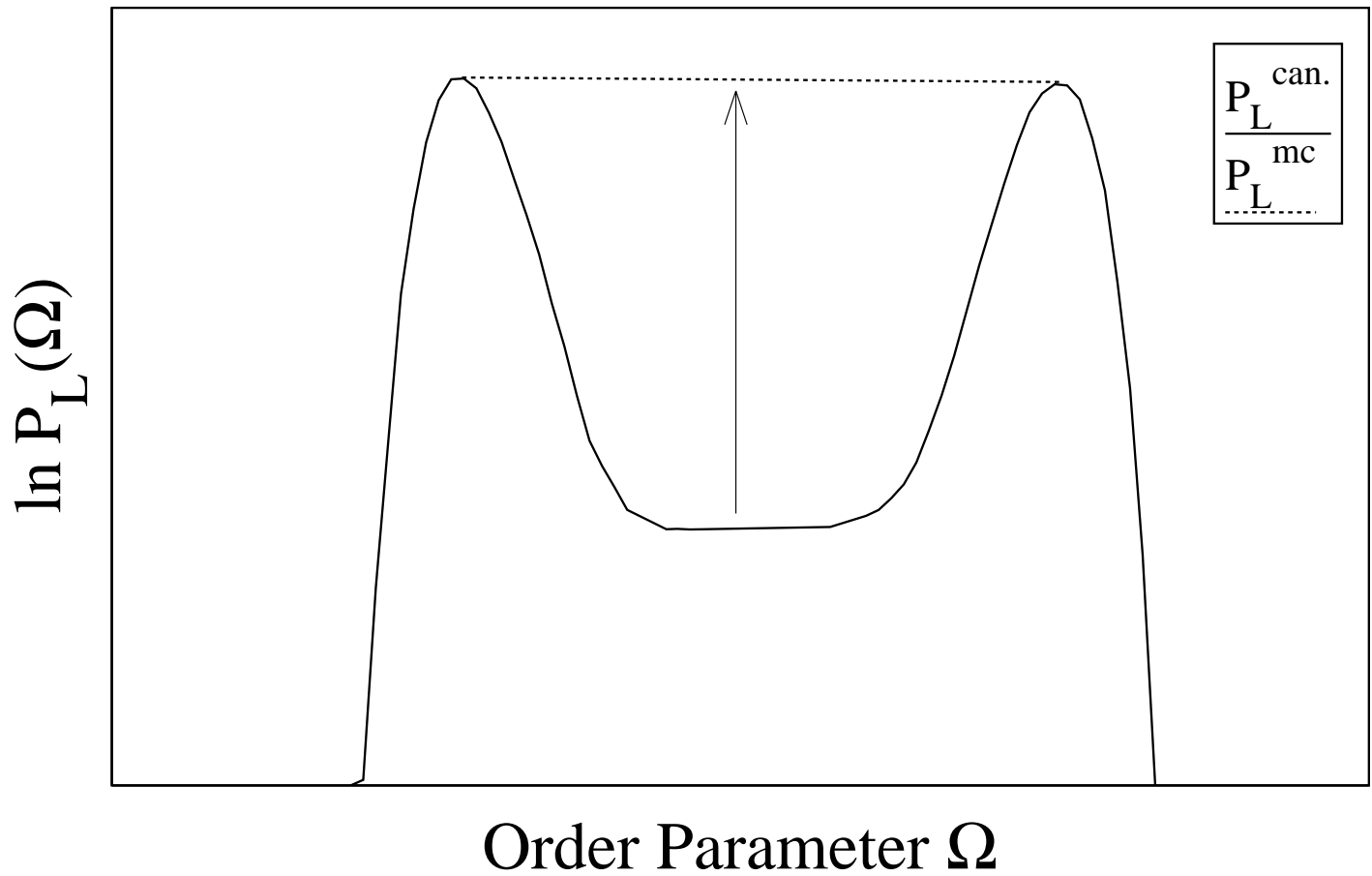
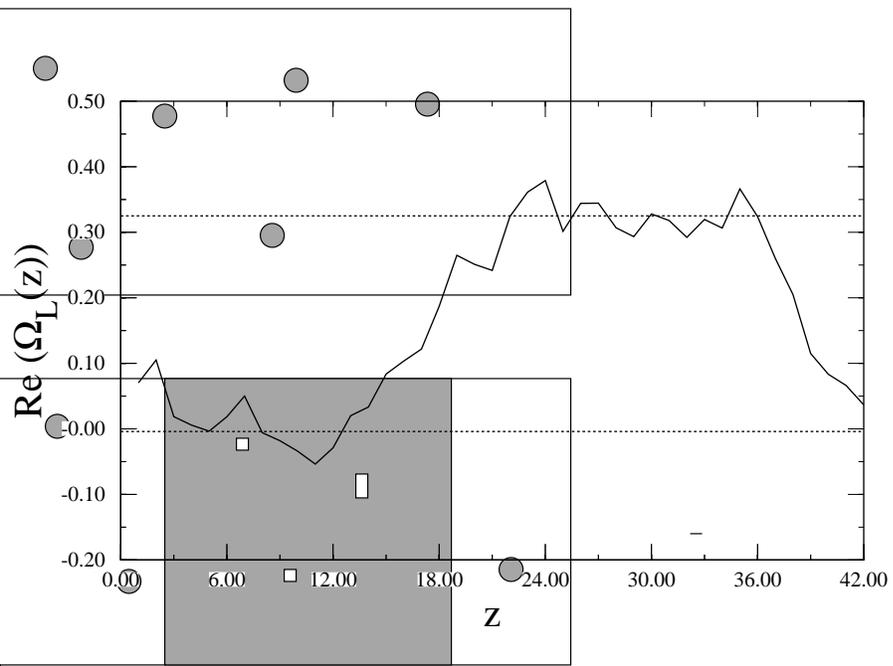
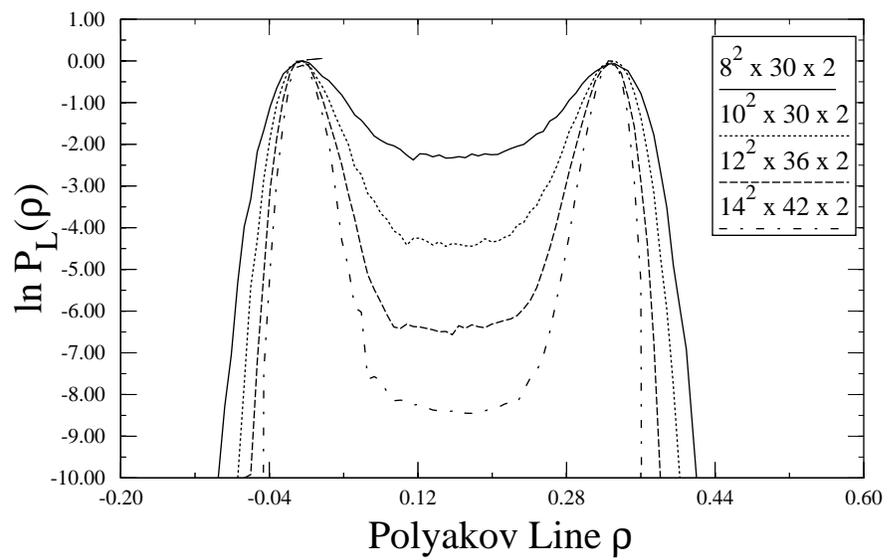
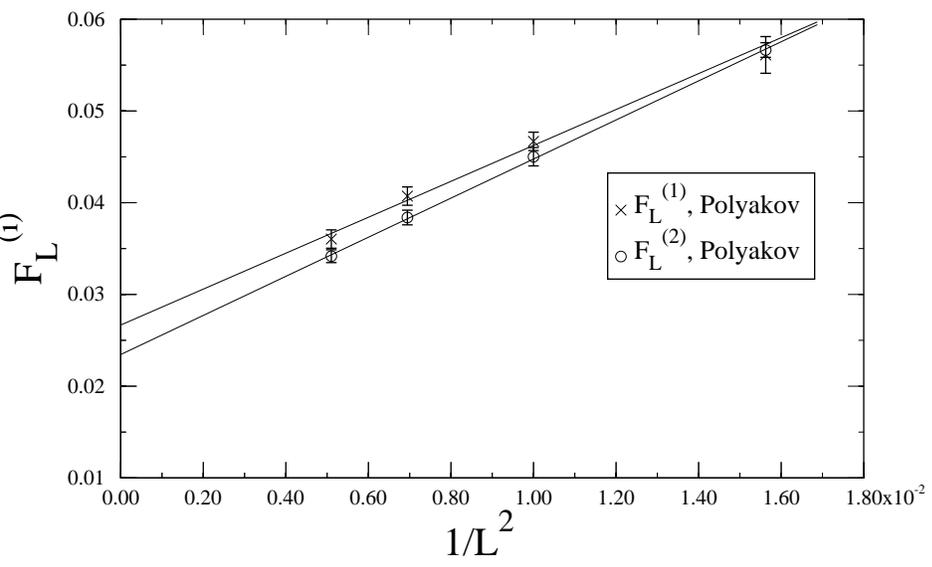
\end{document}